\documentclass[preprint,secnumarabic,nobalancelastpage,nofootinbibt]{revtex4}

\usepackage{graphics}      
\usepackage{graphicx}      
\usepackage{longtable}     
\usepackage{url}           
\usepackage{bm,color}            
\usepackage{amssymb, amsmath, enumerate, theorem, epsfig,subfigure,setspace}
\usepackage{amsfonts}

\begin{document}

\title{Simulating Ising and Potts models and  external fields with non-equilibrium condensates}
  \author{Kirill P. Kalinin$^1$ and Natalia G. Berloff$^{2,1}$ }
\email[correspondence address: ]{N.G.Berloff@damtp.cam.ac.uk}
\affiliation{$^1$Department of Applied Mathematics and Theoretical Physics, University of Cambridge, Cambridge CB3 0WA, United Kingdom}
\affiliation{$^2$Skolkovo Institute of Science and Technology Novaya St., 100, Skolkovo 143025, Russian Federation}

\date{June 2, 2018}

\begin{abstract}{Classical spin models with discrete or continuous degrees of freedom arise in  many studies of complex physical systems. A wide class of hard real-life optimisation problems can be formulated  as a minimisation of a spin Hamiltonian.  Here we show how to simulate the discrete Ising and  Potts models with or without  external fields   using  the physical gain-dissipative platforms with continuous phases such as  lasers and various non-equilibrium Bose-Einstein condensates. The underlying operational principle originates from a combination of resonant and non-resonant pumping. Our results lay grounds for the physical simulations of a broad range of Hamiltonians with complex interactions  that can vary in time and space and with combined symmetries.  }  
\end{abstract}

\maketitle

Understanding  of the phase transitions, structures of the ground states, dynamical behaviour of classical spin systems lies at the core of  studies of complex physical systems.  The interplay
of different types of excitations, couplings, continuous and discrete degrees of freedom
can lead to  complex behaviour that is mostly unaccessible. On the one hand, the majority of such spin problems  are computationally impractical for conventional classical computers for sufficiently large system sizes as their complexity grows exponentially fast with the number of degrees of freedom. On the other hand, experimental studies in solid-state
systems are challenging for  implementing, isolating and controlling such spin model Hamiltonians.

  Beyond the investigation of new physical regimes,
not realisable in condensed matter systems lies another important area of research where finding the ground state configuration of spin models is associated with  classical optimization (NP-hard) problems in vastly different areas such as  vehicle routing and scheduling problems, dynamic analysis of neural networks and financial markets, prediction of new chemical materials and machine learning. 
 Recently various analog Hamiltonian optimisers and quantum devices to simulate classical NP-hard problems have been proposed and realised using  different types of  physical platforms. Regardless of a particular physical system used, the operational principle  of these analog simulators  is substantially similar:  a classical optimisation problem is reformulated as the problem of finding the ground state of a particular spin Hamiltonian with discrete or continuous degrees of freedom that a given simulator can emulate. The discrete NP-hard combinatorial optimization problems such as travel salesman, graph colouring, partitioning, etc. were explicitly mapped into the Ising Hamiltonian \cite{lucas} whereas the continuous quadratic optimization problems such as the phase retrieval is formulated as  finding the global minimum of the XY Hamiltonian \cite{candes11}. The Potts Hamiltonians with more than two discrete states  \cite{potts} appear in many areas  of research such as statistical mechanics including polymer gelation \cite{pottspolymer}, site percolation in the lattice gas \cite{pottslattice}, they related to the free energy of protein folding \cite{pottsProtein} and cellular structural coupling \cite{pottscell}. 

Among the various platforms that aim to  simulate the classical spin Hamiltonians a new subclass of simulators has recently emerged -- the gain-dissipative simulators -- non-equilibrium systems that use a gain process to raise the system above the threshold for a phase transition to a coherent state (via a supercritical Hopf bifurcation) at the global minimum of an associated functional that describes the occupation of this state. Such platforms are mostly optical  such as injection-locked laser systems \cite{yamamoto11},
the networks of optical parametric oscillators, \cite{yamamoto14, yamamoto16a, yamamoto16b,takeda18},  coupled lasers \cite{coupledlaser}, and non-equilibrium condensates such as polariton condensates \cite{NatashaNatMat2017}, and photon condensates \cite{KlaersNatPhotonics2017}. The `spin' in such systems is represented by the phase $\theta_i$ of the complex amplitude $\Psi_i$ that describes the state of the $i-$th laser, oscillator or condensate depending on the nature of the system. We will refer to these states as coherent centres (CCs). The phases of the complex amplitudes take continuous values in $[0,2 \pi)$  and  the system may simulate the XY Hamiltonian $H_{XY}=-\sum J_{ij} \cos(\theta_i-\theta_j)$, as it has been shown in coupled lasers \cite{coupledlaser}, non-degenerate optical parametric oscillators \cite{takeda18}, and non-equilibrium condensates \cite{NatashaNatMat2017,KlaersNatPhotonics2017}. For a general form of the matrix of couplings $J_{ij}$  finding the ground state of such  Hamiltonian  is known to be NP-hard  \cite{ZHANG06};  therefore, any NP problem can be mapped into the XY Hamiltonian but with a polynomial overhead on the number of nodes that can be rather significant  \cite{cubitt16}. Combinatorial optimization problems are usually discrete, so mapping them to a continuous Hamiltonian is not practical. It is important to be able to reduce the overhead by mapping discrete problems into discrete spin Hamiltonians directly. 

The nature of the couplings varies between the physical systems, so in this Letter we focus on non-equilibrium condensates  and coupled lasers. So far in any realisation of a lattice of condensates and lasers the `spins' -- phases of CCs  could take any value between $0$ and $2\pi$. There were various attempts to map the continuous  phases into Ising discrete spins.  For instance, in the injection-locked lasers platform \cite{yamamoto11} and exciton-polariton systems \cite{ohadi17} the phases of the coherent states are projected on  the spin up and down polarisation configurations depending on the occupation of left and right circularly polarised  states; however, there is no evidence that such projection indeed minimises the Ising  Hamiltonian.  
In this Letter we describe the procedure for implementing the discrete Hamiltonians -- the Ising and the Potts Hamiltonians -- in the geometrically coupled non-equilibrium condensates which can be adapted to other gain-dissipative systems. 
The result is a flexible model system which allows us to find the time and pumping-dependent behaviour and interplay of the discrete and continuous order parameters with different symmetry breaking properties.

To realise the XY model using the gain-dissipative simulators one needs to   establish a feedback connection between the gain mechanism, coupling strength and the densities of the CCs  as we recently established \cite{GD-polariton}. The time evolution of the system of $N$ CCs in the gain-dissipative simulator is described by the rate equations
\begin{eqnarray}
 \frac{d \Psi_i}{d t}&=& \Psi_i (\gamma_i-\sigma|\Psi_i|^2) - {\rm i} U|\Psi_i|^2\Psi_i \nonumber\\
 &+& \sum_{j, j\ne i}^N \Delta_{ij}^{\rm inj}K_{ij}\Psi_j + D \xi_i(t), 
\label{ai}\\
\frac{d \gamma_i^{\rm inj}}{dt} &=& \epsilon (\rho_{\rm th} - \rho_i), \quad
\frac{d K_{ij}}{d t} = \hat{\epsilon}( J_{ij}-\Delta_{ij}^{\rm inj}K_{ij} ),\label{gamma2}
\end{eqnarray}   
where $\Psi_i(t)=\sqrt{\rho_i(t)}\exp[{\rm i} \theta_i(t)]$ is a complex amplitude of the $i-$th CC with number density $\rho_i$ and phase $\theta_i$, $\gamma_i=\gamma_i^{\rm inj}  - \gamma_c$ is the effective gain with $\gamma_i^{\rm inj}$ being the pumping rate, $\gamma_c$ and $\sigma$ are rates of  linear and nonlinear dissipation respectively,  $\rho_{\rm th}$ is the specified threshold number density, $\Delta_{ij}^{\rm inj}K_{ij}$ is the coupling strength between $i-$th and $j-$th CCs. For spatially separated condensates $K_{ij}$ depends on  the distance between the condensates, for trapped condensates on the hight of the barrier between them, etc. Dependence of the interaction strengths on the pumping intensity is incorporated into $\Delta_{ij}^{\rm inj}=\gamma_i^{\rm inj} + (1-\delta) \gamma_j^{\rm inj},$ where $\delta$ is an asymmetry parameter. The parameter $\delta=0$ if the interactions are symmetric, as in geometrically coupled condensates that interact via the outflow of the particles from one condensate to another. The parameter $\delta=1,$ if the interactions are unidirectional, so the particles from the $i-$th CC are inserted into  the $j-$th CC, but not the other way around (e.g. as can be achieved in the networks of optical parametric oscillators \cite{yamamoto14}.) The constants $\epsilon$ and $\hat{\epsilon}$ characterize the rates of the density and coupling adjustments respectively:  if the number density (the coupling strength) of the $i-$th CC is below [above] $\rho_{\rm th}$ ($J_{ij}$) it has to be increased [decreased]. For instance, such a feedback can be easily achieved by measuring  photoluminescence of the signal and reconfiguring the spacial light modulator that controls the intensity and configuration of the lattice. The delta-function interactions (pseudo) potential between the particles within each CC has strength $U$ and acts to desynchronise the CCs. Finally,  $\xi_i(t)$ represents the Langevin noise that allows the system to span various phase configurations when approaching the threshold while the diffusion coefficient $D$ disappears at the threshold.
If the pumping rates are driven from below the threshold the fixed points  of Eqs.~(\ref{ai},\ref{gamma2}) coincide with the global minimum of the XY Hamiltonian $H_{XY}$ as we have previously shown \cite{GD-polariton}.

To implement the external fields and the discrete versions of the XY model such as the Ising and the Potts models we need to break the symmetry of Eq.~(\ref{ai}) to phase rotations which can be done by forcing the system parametrically at a frequency $\omega_c$ resonant with the basic frequency of the system at the Hopf bifurcation $\omega_0$. Such resonant forcing has recently been used in combination with a non-resonant pumping in polariton condensates to switch the polarisation states experimentally \cite{prlOhadi16}, where one 
 of the  two condensates was resonantly excited with a narrow linewidth cw diode
laser. For an integer ratio $n=\omega_c/\omega_0$ the Eq.~(\ref{ai}) becomes in analogy with \cite{resonant,resonant2}
\begin{eqnarray}
 \frac{d\Psi_i}{d t}&= &\Psi_i (\gamma_i-\sigma|\Psi_i|^2)- {\rm i} U|\Psi_i|^2\Psi_i + \sum_{j, j\ne i}^N \Delta_{ij}^{\rm inj}K_{ij}\Psi_j\nonumber\\
 &+& h_{ni} \Psi_i^{*(n-1)} + D \xi_i(t).
\label{air}
\end{eqnarray} 
where $h_{ni}$ is the pumping strength of  $i-$th CC at the resonant frequency $n$. In Eq.~(\ref{air}) we use the Madelung transformation $\Psi_i(t)=\sqrt{\rho_i(t)}\exp[{\rm i} \theta_i(t)]$ and separate real and imaginary parts dropping the noise term for simplicity to get 
\begin{eqnarray}
\frac{1}{2}\dot{\rho}_i(t)&=&(\gamma_i -\sigma \rho_i) \rho_i 
+ \sum_{j;j\ne i} \Delta_{ij}^{\rm inj}K_{ij} {\sqrt{\rho_i\rho_j}}\cos\theta_{ij}\nonumber \\ &+& h_{ni} \rho_i^{\frac{n}{2}}\cos (n \theta_i),\label{rho}\\
\dot{\theta}_i(t)&=&-  U\rho_i-\sum_{j;j\ne i} \Delta_{ij}^{\rm inj}K_{ij} {\frac{\sqrt{\rho_j}}{\sqrt{\rho_i}}} \sin\theta_{ij} \nonumber\\
&- &h_{ni}\rho_i^{\frac{n}{2}-1}\sin(n \theta_i),\label{theta}
\end{eqnarray}
where $\theta_{ij}=\theta_i-\theta_j$. The fixed point of the dynamical system given by   the Eqs.~(\ref{rho},\ref{theta},\ref{gamma2}) satisfies
\begin{equation}
\rho_i=\rho_{\rm th} = [ \gamma_i + \sum_{j;j\ne i} J_{ij} \cos\theta_{ij} +h_{ni} \rho_{\rm th}^{\frac{n}{2}-1}\cos (n \theta_i)]/\sigma.
\end{equation}
Since for each CC we choose the smallest $\gamma_i$ by raising it from below then  at the threshold the global minimum of 
\begin{equation}
H=-\sum_{i=1}^N\sum_{j=1;j\ne i}^N J_{ij} \cos\theta_{ij} -\rho_{\rm th}^{\frac{n}{2}-1}\sum_{i=1}^N h_{ni} \cos (n \theta_i) \label{h}
\end{equation}
  is achieved while Eq.~(\ref{theta}) describes the gradient decent to that minimum. By taking the resonance $n=1$ we introduce the effective external "magnetic" field ${\bf F}=\{h_{1i}/\sqrt{\rho_{\rm th}}\}$ into the model. 
For $n>1$ the forcing term in Eq.~(\ref{air}) reduces the invariance to a global phase shift to a discrete symmetry $\theta_i=2 \pi i/n $ and for a sufficiently large $h_{ni} \rho_{\rm th}^{\frac{n}{2}-1}>\sum_j |J_{ij}|$ introduces the penalty term in the Hamiltonian $H$ for the deviation of phases from the discrete values $2 \pi i/n.$ For $n=2$ and  a uniform strength of the resonant pumping $h_2=h_{2i},$  Eqs.~(\ref{air},\ref{gamma2}) realise the global minimum of Eq.~(\ref{h}) with $\theta_i$ restricted to $0$ or $\pi$, and therefore, the global minimum of the Ising Hamiltonian, whereas for $n>2$ and ${h}_n=h_{ni}\rho_{\rm th}^{\frac{n}{2}-1}$ Eqs.~(\ref{air},\ref{gamma2})  realise the Potts Hamiltonian with $\theta_i=2 \pi i/n$ \cite{potts}. 

\begin{figure}[t!]
\centering
  \includegraphics[width=8.6cm]{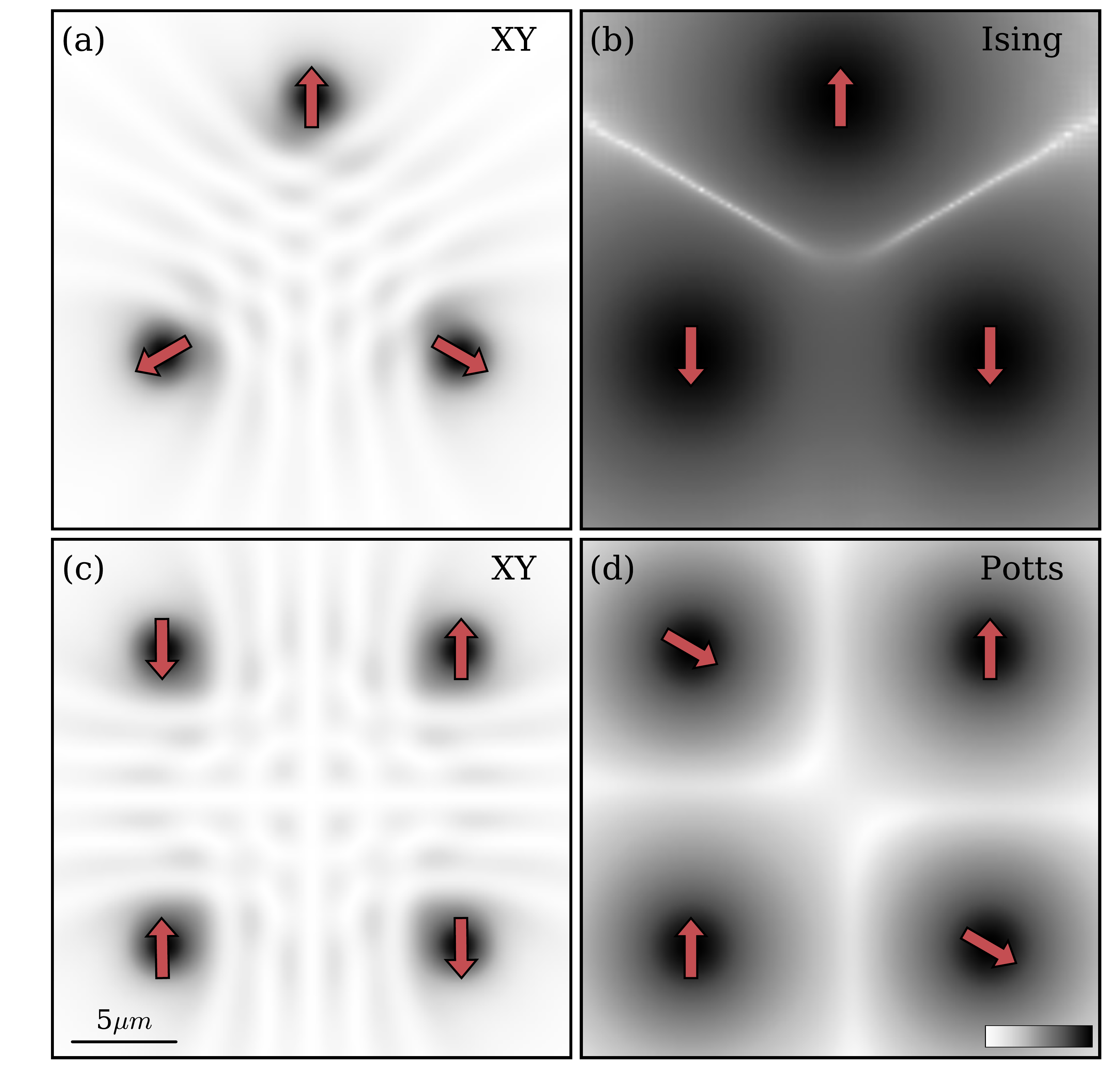}
\caption{Contour plots of the density $|\psi({\bf r})|^2$ of the polariton condensates with the nearest neighbours anti-ferromagnetically coupled  realising the ground state of (a,c) the XY  (b)  the Ising and (d) the 3-state Potts Hamiltonians at the condensation threshold. The densities for the Ising and Potts Hamiltonians are shown in log scale to emphasise the standing matter waves  between the condensates, and, therefore, the phase differences. The densities are obtained by numerical integration of Eqs.~(\ref{Initial Reservoir equation}) with the parameters given in \cite{numer}.  }
 \label{Figure1}
\end{figure}

The Ising Hamiltonian has been previously realised in optical networks, most notably  in the degenerate optical parametric oscillators (DOPOs) \cite{dopo}, where the phase projection to $0$ and $\pi$ is due to the  second order nonlinear crystal placed in an optical cavity. However, only resonant forcing is present there, which limits the range of couplings. Our proposal is the first scheme to realise the Ising Hamiltonian   using   lattices of  non-equilibrium Bose-Einstein condensates such as polariton or photon condensates and  the first proposal to  realise  the Potts model as well as   the combination of continuous and discrete spins in any  optical system. In other systems, for instance in  ultracold optical lattices   \cite{opticallattices} the combination of XY and Ising models has been realised  using tunable artificial gauge fields, but such systems have no means to address the ground state of the spin Hamiltonian.

To simulate the behaviour of the gain-dissipative system one can use Eqs.~(\ref{air},\ref{gamma2}); however, since simplifying assumptions were made to arrive at Eq.~(\ref{air}) it is important to verify that a real system shows the transition  from achieving the minimum of the continuous (XY) model to  the minimum of the discrete  model when the resonant forcing  is introduced. Therefore,  we consider, as an illustrative example,  the exciton-polariton system modelled by the complex Ginzburg-Landau equation coupled to the rate equation describing the  hot exciton reservoir \cite{carusotto, goveq}. In such a system polaritons are excited non-resonantly with a single-mode continuous wave laser using  a spatial light modulator  to generate patterns of laser spots on the sample surface as in our experimental work \cite{NatashaNatMat2017} and with the same  parameters.  As we established in \cite{NatashaNatMat2017}, such a model reliably represents the behaviour of polariton lattices.  The resonant drive is achieved by another single-mode continuous wave laser tuned to the multiple of the frequency of the system at the Hopf biburcation \cite{hopf, resonant}. The resulting equations read
\begin{eqnarray}	
	i \hbar  \frac{\partial \psi}{\partial t} &=& - \frac{\hbar^2}{2m}  \left(1 - i \eta_d {\cal R} \right) \nabla^2\psi + U_0 |\psi|^2 \psi+
	\hbar g_R {\cal R} \psi  \nonumber \\
	  &+&i\hbar \biggl(R_R {\cal R} - \gamma_C \biggr) \psi +  P_{n}({\bf r},t)\psi^{*(n-1)}, \label{Initial GL equation}\\
	  \frac{\partial \cal R}{\partial t} &=&  - \left( \gamma_R + R_R |\psi|^2 \right) {\cal R} + P_{0}({\bf r}) ,
	\label{Initial Reservoir equation}
\end{eqnarray}
where $\psi({\bf r},t)$ is the condensate wavefunction, ${\cal R}({\bf r},t)$ is the density profile of the hot exciton reservoir, $m$ is the polariton effective mass, $U_0$ and  $g_R$ are the strengths of  the effective polariton-polariton interaction  and the blue-shift due to interactions with non-condensed particles, respectively, $R_R$ is the rate at which the exciton reservoir feeds the condensate, $\gamma_C$ is  the decay rates of condensed polaritons,  $\gamma_R$ is the rate of the redistribution of hot excitons in the  reservoir,  $\eta_d$ is the rate of the energy relaxation, and $P_0({\bf r},t)$ is the rate of non-resonant pumping  into the exciton reservoir \cite{numer}. The last term on the right-hand side of Eq.~(\ref{Initial GL equation}) is the resonant forcing with resonance $n$. The polariton lattice of $N$ condensates at the positions ${\bf r}={\bf r}_i$ is formed by taking the  non-resonant pumping profile as $P_{0}({\bf r},t)=\sum_{i=1}^N f_i (t) p(|{\bf r}-{\bf r}_i|)$, where $p(r)=\exp(-\alpha r^2)$, $\alpha$ characterizes  the inverse width of the incoherent pumping profile and $f_i$ describes the strength of the pumping centred at the position ${\bf r}={\bf r}_i$. The resonant pumping profile $P_{n}({\bf r},t)= \sum_{i=1}^N \tilde h_{ni} p(|{\bf r}-{\bf r}_i|), n> 0$ follows the lattice spatial profile but with different pumping intensities $\tilde h_{ni}\ll f_i$. As we have previously shown   \cite{GD-polariton} the spatial degrees of freedom of Eqs. ~(\ref{Initial GL equation}, \ref{Initial Reservoir equation}) without the resonant terms can be integrated out to yield the rate equations on the complex amplitudes of the CCs leading to Eqs.~(\ref{ai}). Similarly, Eqs. ~(\ref{Initial GL equation}, \ref{Initial Reservoir equation}) with the resonant forcing yield Eqs.~(\ref{air}) with $ h_{ni}=\tilde h_{ni}  Re[\int  p(|{\bf r}-{\bf r}_i|) \phi^{*n}(|{\bf r}-{\bf r}_i|)\, d{\bf r}]/\int |\phi(|{\bf r}-{\bf r}_i|)|^2\, d{\bf r},$ where $\phi(r)$ is the wavefunction of a single condensate pumped with $p(r)$.

To illustrate how the polariton lattice minimizes the discrete Ising or Potts models with or without external fields we study the behaviour of  the unit polariton lattice cells when they are subjected to the effect of the resonant forcing. In what follows we explore the behaviour of the system without density and coupling adjustments described by Eqs.~(\ref{gamma2}) to see how the introduction of the resonant forcing changes the couplings.   The effect of such adjustments is elucidated elsewhere \cite{GD-algorithm}.
First, we consider a simple lattice of three condensates arranged at the corners of the equilaterial triangle \cite{keelingBerloffarxiv, baumbergGeometrical, ohadi16} coupled antiferromagnetically with $J=J_{ij}<0$. Without the resonant forcing, $P_{n}=0,n>0,$  the phases arrange themselves with $2 \pi/3$ phase differences to minimize $H_{XY}=-J(\cos\theta_{12} + \cos\theta_{23} +\cos\theta_{31})$ as Fig.~\ref{Figure1}a illustrates. This agrees with the experimental findings \cite{ohadi16}. When the resonant forcing  is introduced with $n=2$  the system described by  Eqs. ~(\ref{Initial GL equation}, \ref{Initial Reservoir equation}) finds the global minimum of the Ising Hamiltonian $H_{I}=-\tilde{J}(s_1 s_2 +  s_2 s_3 + s_1 s_3),s_i=\cos \theta_i=\pm1$, for $\tilde{J}<0, \tilde{J}\ne J$. The system is frustrated: spins $0,\pi,\pi$ or $0,0,\pi$ are realised as  Fig.~\ref{Figure1}b depicts.

To illustrate the transition from solving the XY model to the Potts model we consider four condensates arranged at the corners of a square  with antiferromagnetic coupling between the nearest neighbours and ferromagnetic coupling along the diagonal.  Figure \ref{Figure1}c shows the solution of Eqs. ~(\ref{Initial GL equation}, \ref{Initial Reservoir equation}) without the resonant terms ($P_n=0, n>0$). Four condensates  realise the global minimum of the XY model with $0,\pi,0,\pi$ phase differences  as been also observed in experiments \cite{baumbergGeometrical,NatashaNatMat2017}. The same configuration would result from the Ising model, but the 3-state  Potts model with $\theta_i$ restricted to $0, 2 \pi/3$ and $4 \pi/3$   is minimized by $0,z,0,z,0$ configurations where $z=2 \pi/3$ or $z=4 \pi/3.$ This is what we observe implementing $n=3$ resonant forcing in Eqs. ~(\ref{Initial GL equation}, \ref{Initial Reservoir equation}) as Fig.~\ref{Figure1}d illustrates. 

\begin{figure}[t!]
\centering
  \includegraphics[width=8.6cm]{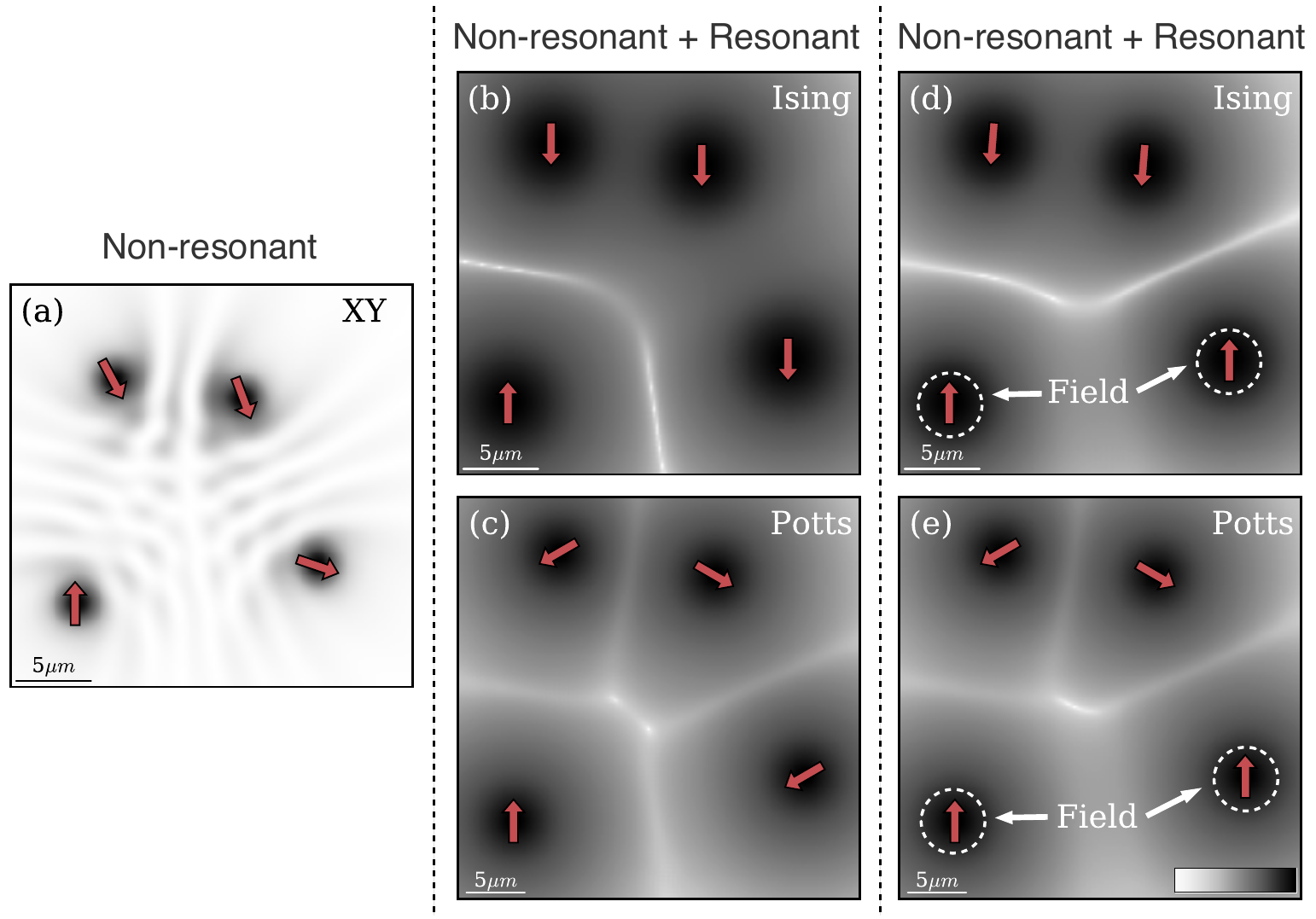}
\caption{Contour plots of the density $|\psi({\bf r})|^2$ of the polariton condensates  in the corners of quadraterial realising the ground state of (a) the XY  (b)  the Ising and (c) the 3-state Potts Hamiltonians without the external fields; (d) the Ising and (e) the 3-state Potts Hamiltonians with the external fields forcing the bottom condensates to acquire phase $\theta_i$. The densities for the Ising and Potts Hamiltonians are shown in log scale to emphasise the standing matter waves  between the condensates, and, therefore, the phase differences. The densities are obtained by numerical integration of Eqs.~(\ref{Initial Reservoir equation}) with the parameters given in \cite{numer}.  }
 \label{Figure2}
\end{figure}

Finally, we combine two resonant forcing terms: the resonance $n= 1$ and either resonance $n=2$ or $n=3$   in Eq.~(\ref{Initial GL equation}) to combine the effect of an external "magnetic" field in the Ising or  3-state Potts models respectively. We take  $\tilde h_{11}=\tilde h_{12} > {\rm max} |J_{ij}|$ and $\tilde h_{13}=\tilde h_{14}=0$. Such external field  penalises the objection function if the phases of the bottom two condensates on Fig.~\ref{Figure2} (b-c) are not zeros and leads to the phase configurations as shown in Fig.~\ref{Figure2} (d-e).

In conclusion,  we formulated an implementation of the discrete spin Hamiltonians such as the Ising and Potts models with or without the external fields in a lattice of non-equilibrium Bose-Einstein condensates or coupled lasers.  We show that if a resonant pumping is employed together with a non-resonant  excitation then depending on the nature of the resonance the phases of the condensates or lasers in a  lattice take discrete values and at the threshold implement the minimum of the corresponding discrete spin Hamiltonian. This approach opens  rather exciting possibilities for  simulating complex physical systems, solving  combinatorial  optimisation problems and developing new computational algorithms. Such  lattices can  be used to study the dynamical phase transitions and novel regimes of spin systems and the proposed approach offers a unique platform with  wealth of controllable and tuneable parameters. The strengths of the resonant and non-resonant pumping can  be varied in both time and space, between lattice sites, and even the co-existing lattices with different  `magnetic' properties can be easily simulated. Our results allow the investigation of new physical regimes, not before realisable in condensed matter systems.

\section*{Acknowledgements}
The authors acknowledge financial support from the NGP MIT-Skoltech. K.P.K. acknowledges the financial support from Cambridge Trust and EPSRC.


\begin{thebibliography}{00}
\bibitem{lucas} Lucas, A. Ising formulations of many NP problems. {\it Front. Phys.} {\bf 2}, 5 (2014).
%
\bibitem{candes11} Candes, E. J., Eldar, Y. C., Strohmer, T., \& Voroninski, V.  Phase retrieval via matrix completion. \textit{SIAM review} \textbf{57(2),} 225-251 (2015).
%
\bibitem{potts} Wu, F. Y. The potts model. \textit{Rev. Mod. Phys.} 5\textbf{4(1),} 235 (1982).
%
\bibitem{pottspolymer} Coniglio, A., Stanley, H. E., \& Klein, W.  Site-bond correlated-percolation problem: a statistical mechanical model of polymer gelation. \textit{Phys. Rev. Lett.} \textbf{42(8),} 518 (1979).
%
\bibitem{pottslattice} Murata, K. K.  Hamiltonian formulation of site percolation in a lattice gas. \textit{J. of Phys. A: Math.  Gen.} \textbf{12(1),} 81 (1979).
%
\bibitem{pottsProtein} Morcos, F., Schafer, N. P., Cheng, R. R., Onuchic, J. N., \& Wolynes, P. G. Coevolutionary information, protein folding landscapes, and the thermodynamics of natural selection. \textit{Proc. Nat. Acad. Sci.} \textbf{111(34),} 12408-12413 (2014). 
%
\bibitem{pottscell} Graner, F., \& Glazier, J. A.  Simulation of biological cell sorting using a two-dimensional extended Potts model. \textit{Phys. Rev. Lett.} \textbf{69(13)}, (1992).
%
\bibitem{yamamoto11} Utsunomiya, S., Takata, K. \& Yamamoto, Y. Mapping of Ising models onto injection-locked laser systems. \textit{Opt. Express} \textbf{19}, 18091 (2011).
%
\bibitem{yamamoto14} Marandi, A., Wang, Z., Takata, K., Byer, R.L. \& Yamamoto, Y. Network of time-multiplexed optical parametric oscillators as a coherent Ising machine. \textit{Nat. Phot.} \textbf{8}, 937-942 (2014).

\bibitem{yamamoto16a}  Inagaki, T.  \textit{et al}. Large-scale Ising spin network based on degenerate optical parametric oscillators. \textit{ Nat. Phot.} {\bf 10},415-419 (2016).

\bibitem{yamamoto16b} McMahon, P.L. \textit{et al}. A fully programmable 100-spin coherent Ising machine with all-to-all connections. \textit{Science}, {\bf 354} 614-617 (2016).
%
\bibitem{takeda18} Takeda, Y. \textit{et al}. Boltzmann sampling for an XY model using a non-degenerate optical parametric oscillator network. \textit{Quan. Sci. Tech.} \textbf{3(1),} 014004 (2017). 
%
\bibitem{coupledlaser} Nixon, M., Ronen, E., Friesem, A. A. \& Davidson, N. Observing geometric frustration with thousands of coupled lasers. \textit{Phys. Rev. Lett.} \textbf{110}, 184102 (2013).
%
\bibitem{NatashaNatMat2017}  Berloff, N. G. \textit{et al}. Realizing the classical XY Hamiltonian in polariton simulators. \textit{Nat. Mat.} \textbf{16(11)}, 1120 (2017).
%
%
\bibitem{KlaersNatPhotonics2017} Dung, D. \textit{et al}. Variable potentials for thermalized light and coupled condensates. \textit{Nat. Phot.} \textbf{11(9)}, 565 (2017).
%

%
\bibitem{ZHANG06} Zhang, S. \& Huang, Y. Complex quadratic optimization and semidefinite 
programming. \textit{SIAM J. Optim.} {\bf 16}, 871 (2006).
%
\bibitem{cubitt16} De las Cuevas, G., \& Cubitt, T. S. Simple universal models capture all classical spin physics. \textit{Science} \textbf{351(6278),} 1180-1183 (2016). 

\bibitem{ohadi17} Ohadi, H. \textit{et al.} Spin order and phase transitions in chains of polariton condensates. {\it Phys. Rev. Letts.}, {\bf 116}, 067401 (2017).

\bibitem{GD-polariton} Kalinin, K. P. \& Berloff, N. G. Gain-dissipative simulators for large-scale hard classical optimisation. arXiv:1805.01371 (2018). 

\bibitem{prlOhadi16} Ohadi, H. {\it et al.} Tunable Magnetic Alignment between Trapped Exciton-Polariton Condensates. {\it Phys. Rev. Letts.} {\bf 116}, 106403 (2016).

\bibitem{resonant} Coullet, P. \& K Emilsson, K. Strong resonances of spatially distributed oscillators: a laboratory to study patterns and defects. {\it Physica D: Non. Phen.} {\bf 61,}  119 (1992).

\bibitem{resonant2} Yochelis. A., Elphick, C., Hagberg, A., \& Meron, E. Two-phase resonant patterns in forced oscillatory systems: boundaries, mechanisms and forms. {\it  Physica D: Non. Phen.} {\bf 199,} 201 (2004).

%
\bibitem{dopo} Wang, Z., Marandi, A., Wen, K., Byer, R. L., \& Yamamoto, Y. Coherent Ising machine based on degenerate optical parametric oscillators. \textit{Phys. Rev. A} \textbf{88(6),} 063853 (2013). 

\bibitem{opticallattices} Struck, J. \textit{et al.} Engineering Ising-XY spin-models in a triangular lattice using tunable artificial gauge fields. \textit{Nat. Phys.} \textbf{9(11),} 738 (2013). 

\bibitem{goveq}  Keeling, J. \& Berloff, N. G. Spontaneous rotating vortex lattices in a pumped decaying condensate. \textit{Phys. rev. lett.} \textbf{100(25),} 250401 (2008). 

\bibitem{carusotto} Wouters, M. \& Carusotto, I. Excitations in a nonequilibrium Bose-Einstein condensate of exciton polaritons. \textit{Phys. Rev. Lett.} \textbf{99}, 140402 (2007).

\bibitem{hopf} Kuramoto, Y. Chemical oscillations, waves, and turbulence (Dover, New York, 2012).

\bibitem{GD-algorithm} Kalinin, K.P. \& Berloff, N. G. Global optimization of spin Hamiltonians with gain-dissipative systems. in preparation (2018).

\bibitem{numer} The numerical evolution of Eqs. (\ref{Initial Reservoir equation})  were performed using the 4th-order Runge-Kutta integration in time and 4th order finite difference scheme in space starting with many random complex noise configurations. The parameters were the same as in our previous work \cite{NatashaNatMat2017}. In addition, the following parameters were used to simulate the resonant pumping: $P_2({\bf r},t) = 0.5( \tanh (6 t / t_{max} - 3) + 1) \sum_{i=1}^N p(|{\bf r - r_i}|)$ , $P_3({\bf r},t) = 0.25( \tanh (6 t / t_{max} - 3) + 1) \sum_{i=1}^N p(|{\bf r - r_i}|)$, and to simulate the field $P_1({\bf r},t) = 0.25( \tanh (6 t / t_{max} - 3) + 1) (p(|{\bf r - r_1}|) + p(|{\bf r - r_2}|))$, where $t_{max}\approx 100$ is the time when a steady state is achieved.

\bibitem{keelingBerloffarxiv} Keeling, J., \& Berloff, N. G. Controllable half-vortex lattices in an incoherently pumped polariton condensate.  arXiv:1102.5302 (2011).

\bibitem{baumbergGeometrical} Tosi, G. \textit{et al.} Geometrically locked vortex lattices in semiconductor quantum fluids. \textit{Nat. Comm.} \textbf{3}, 1243 (2012).

\bibitem{ohadi16} Ohadi, H., Gregory, R. L., Freegarde, T., Rubo, Y. G., Kavokin, A. V., Berloff, N. G., \& Lagoudakis, P. G. Nontrivial phase coupling in polariton multiplets. \textit{Phys. Rev. X} \textbf{6(3),} 031032 (2016). 


\end{thebibliography}
\end{document}